\documentclass[]{aa}
\usepackage{graphicx}
\usepackage{setspace}
\usepackage{natbib}
\bibpunct{(}{)}{;}{a}{}{,}
\begin{document}

\def\etal{{\it et\thinspace al.}\ }
\def\cm3{cm$^{-3}$\ }
\def\mic {$\mu$m\ }


\title{Atomic data from the Iron Project. LXIV. 
Radiative transition rates and collision strengths for 
\ion{Ca}{ii}
\thanks{The atomic data from this work, including energy levels, A-values,
 and effective collision strengths, is available in electronic form
 at the CDS via anonymous ftp to cdsarc.u-strasbg.fr (130.79.128.5)
 or via http://cdsweb.u-strasbg.fr/cgi-bin/qcat?J/A+A/.
}} 

\author{ M. Mel\'endez\inst{1}\inst{2}
 \and  M.A. Bautista\inst{3} 
 \and N.R. Badnell\inst{4}}

 \institute{Institute for Astrophysics and Computational 
Sciences, Department of Physics, The Catholic University of America,
 Washington, DC 20064; 07melendez@cua.edu \and Exploration of the Universe Division, 
Code 667, NASA's Goddard Space Flight Center, Greenbelt, MD 20771 \and Centro de F\'{\i}sica, Instituto Venezolano de 
Investigaciones Cient\'{\i}ficas (IVIC), PO Box 21827, Caracas 1020A,
Venezuela; bautista@kant.ivic.ve \and Department of Physics, University of Strathclyde, Glasgow G4 0NG}

\date{Received date/ accepted date}


\abstract{}
{This work reports radiative transition
rates and electron impact excitation rate coefficients for levels of the
n= 3, 4, 5, 6, 7, 8  configurations of \ion{Ca}{ii}.}{The radiative data were  computed using
the Thomas-Fermi-Dirac central potential method in the frozen core
 approximation and includes the polarization  interaction between the
 valence electron and the core using a model potential.
This method allows for configuration interactions (CI) and relativistic effects
in the Breit-Pauli formalism.
Collision strengths in LS-coupling were calculated in the close coupling approximation
with the R-matrix
method. Then, fine structure collision strengths were obtained
by means of the intermediate-coupling frame transformation (ICFT) method
which accounts for spin-orbit coupling effects.
}{ We present extensive comparisons with the most recent calculations 
 and measurements for \ion{Ca}{ii} as well as a comparison between the core polarization results and the 
``unpolarized" values. 
We find that core polarization affects the computed lifetimes
by up to 20\%. Our results are in very close agreement with recent measurements for the 
  lifetimes of 
 metastable levels. 
The present collision strengths were integrated over
a Maxwellian distribution of electron
energies and the resulting effective collision strengths are given for a
wide range of temperatures.
Our effective collision 
 strengths for the resonance transitions are within $\sim$11\% from 
previous values derived from experimental measurements, but disagree with
latter computations using the distorted wave approximation.
}
{}
\keywords{atomic data - atomic processes - line: formation - 
stars: Eta Carinae - Active Galactic Nuclei}

\titlerunning{Atomic data from the Iron Project LXIV}
\authorrunning{M. Mel\'endez et al.}

\maketitle

\section{Introduction}

\ion{Ca}{ii} plays a prominent role in astrophysics. The so-called $H$ and $K$ 
lines of this ion are important probes of solar and stellar 
chromospheres \citep{2006PASP..118..617R}. In the red spectra
 of Active Galactic Nuclei (AGN) the infrared triplet of \ion{Ca}{ii} in emission
 has been used to investigate the correlations with optical \ion{Fe}{ii} \citep{1989A&A...208...47J} and their implications
on the physical conditions of the emitting gas \citep{1989ApJ...347..656F}
.[\ion{Ca}{ii}] optical emission lines together with the infrared [\ion{Fe}{ii}]
are often used as probe of dust content of AGNs \citep{1999ASPC..175..353S}.

\ion{Ca}{ii} has been addressed by numerous theoretical and experimental 
groups. The lifetimes 
$\tau$ for the 4p$~^2{\rm P^o}$ and 3d$~^2$D levels have been measured with 
high precision \citep{1993PhRvL..70.3213J, 2005PhRvA..71c2504K}. Various 
theoretical methods have been used in trying to match these experimental values. The most 
recent calculations of  \cite{1995PhRvA..51.1723L}
 using the Brueckner approximation with third-order correction agree 
within $\sim$1\% with the experimental lifetime 
for the 4p$~^2{\rm P^o}$ levels but are $\sim$11\% too small for the 3d$~^2$D 
metastable levels. The calculations of \cite{2005PhRvA..71c2504K}, 
using a relativistic all-order method which sums infinite sets of many-body 
perturbation theory terms, agree 
within $\sim$0.3\% with the experimental lifetimes of the 
$3d~^2$D metastable levels of \ion{Ca}{ii}, but offer no data for other
levels. \cite{1991PhRvA..44.1531G} computed lifetimes using relativistic many-body perturbation theory 
that agree within $\sim$2\% with experimental values for the $4p~^2{\rm P^o}$ levels 
and within $\sim$6\% for the metastable states. 
In the present work we use the Thomas-Fermi-Dirac central potential with
core polarization interaction to provide a complete set of
accurate A-values for allowed and forbidden transitions
to be used in modeling \ion{Ca}{ii} spectra. 
  
Various calculations of collision strengths have been performed for the 
resonance transitions in \ion{Ca}{ii} 
\citep[see][]{ 1991PhRvA..44.7343Z,1981JPhB...14.4149C,1978JPhB...11.1303K,1970JPhB....3..952S}.
\cite{1977ApL....19...11O} derived effective collision
strengths from experimental cross sections of the resonance $K$ and $H$ lines
of \ion{Ca}{ii} at 3934 and 3968 \AA \ \ by \cite{1973PhRvA...8.2304T}. 
Later, \cite{1975SJETP..42..989Z} published cross sections for
exciting 5s and 4d levels from the ground state, which
are important to estimate the contribution to the 4p level by cascade.
 \cite{1988PhRvA..38.3339M} presented a detailed study of the electron-impact excitation of the (4s-4p) transitions using the close-coupling  approximation 
including a polarization potential. 
More recently \cite{1995A&A...300..627B} used a non-exchange distorted wave approximation including the lowest 7 \ion{Ca}{ii}
terms. 

The IRON Project is an international enterprise devoted to the
computation of accurate atomic data for the iron group elements \citep{1993A&A...279..298H}.  A complete list of publications from this project can be found at
http://www.usm.uni-muenchen.de/people/ip/papers/papers.html.
Within this project
we have been systematically working
on the  data for the low ionization stages of iron peak elements, e.g. radiative and  
collisional rates for Fe{\sc\,i--iv} \citep{1998RMxAC...7..163B}, Ni{\sc\,ii} \citep{bau04}, 
Ni{\sc\,iii} \citep{bau01}, \ion{Ni}{iv} \citep{2005A&A...436.1123M}.
The objective of the present work is to provide accurate and complete
atomic data for a detailed spectral modeling of \ion{Ca}{ii}. Such a model
should be large enough to account for various processes such as 
collisional excitation including cascades from high levels, fluorescence
by line and continuum radiation, and line optical depth effects.

\section{Atomic data}

\subsection{Atomic structure calculations}
We use the atomic structure code AUTOSTRUCTURE \citep{1986JPhB...19.3827B,1997JPhB...30....1B} 
to reproduce the structure of 
the \ion{Ca}{ii} ion. This code is based on the program SUPERSTRUCTURE originally 
developed by \cite{1974CoPhC...8..270E}, but incorporates various improvements and new capabilities like
two-body non-fine-structure operators of the Breit-Pauli Hamiltonian and 
polarization model potentials. In this approach, 
 the wave functions are written as configuration interaction expansions of the type:
\begin{equation}
\psi_i = \sum_j \phi_jc_{ji}, 
 \label{nueve}
\end{equation}
where the coefficients $c_{ji}$ are determined by diagonalization of 
 $\langle \psi_i \mid H \mid \psi_i \rangle$.
Here $H$ is the
 Hamiltonian and the  basic functions $\phi_j$ are constructed from one-electron orbitals generated using the Thomas-Fermi-Dirac model
potential \citep{1969JPhB....2.1028E}, including $\lambda_{nl}$ scaling
parameters which are optimized by minimizing a weighted sum of energies. 
The basic list of configurations and scaling parameters used in this 
work are listed in Table 1. 

\begin{table}[ht]
\caption[]{Configuration  expansion for \ion{Ca}{ii}, and scaling parameters $\lambda_{nl}$  for each
orbital of the configurations ${\rm 1s^22s^22p^63s^23p^6}$$nl$ in the 
Thomas-Fermi-Dirac potential} 
\begin{flushleft}
\begin{tabular}{l}
\hline \noalign{\smallskip}
$nl$ Configurations\\
\hline \noalign{\smallskip}
3d,
4s, 
4p,        
4d,
4f,
5s,
5p,
5d,
5f,  
5g,
6s,
6p,
6d,
6f, \\
6g,  
7s,
7d,
7f,  
7g,
8s,
8d,
8f,
8g\\

\hline
$\lambda_{nl}$ \\1s: 1.43880, 2s: 1.11310, 2p: 1.05670, 
                 3s: 1.10580, \\
3p: 1.09850, 3d: 1.07950,
             4s: 1.08770, 4p: 1.07730, \\
4d: 1.07690, 
                4f: 1.05000, 5s: 1.08510, 5p: 1.07690, \\
                5d: 1.07660, 
5f: 1.04950, 5g: 1.01940, 
                6s: 1.08480, \\
6p: 1.07820, 
6d: 1.07690, 
                6f: 1.04960, 6g: 1.01920, \\
7s: 1.08510,  
                7d: 1.07760, 7f: 1.05010, 7g: 1.01930, \\
                8s: 1.08590, 
8d: 1.07860, 8f: 1.05120,  
                8g: 1.01950  \\
\noalign{\smallskip}
            \hline
         \end{tabular}
\end{flushleft}
\end{table}

Relativistic effects are included in the calculation by means of the Breit-Pauli operators in the form: 

\begin{equation}
H = H_{\rm nr}+H_{\rm bp},
\label{uno}
\end{equation}
where $H_{\rm nr}$ is the usual non-relativistic Hamiltonian and $H_{\rm bp}$ is the 
Breit-Pauli perturbation, which includes one- and two-body operators 
\citep{1970JPhB....3.1571J,1971JPhB....4.1422J, 1974CoPhC...8..270E}.

\subsection{Model potential}
 In order to obtain  accurate orbitals in our  multiconfiguration frozen-core approximation we include 
the polarization interaction between the valence electron and the core in a model potential. We used a model 
potential $V_{pol}$ of the form described by \cite{1976JPhB....9.2983N};

\begin{equation}
V_{\rm pol}(r,\rho)= -\frac{\alpha_{\rm d}}{r^4}\left[ 1-\exp(-r/ \rho)^6 \right],
\label{vpol}
\end{equation}
where $\alpha_{\rm d}$ is the static dipole core polarizability of the ion 
\ion{Ca}{iii} and $\rho$ is adjusted empirically to yield good agreement
with experimental energies. \cite{waller1926}  using the 
``non-penetrating" orbitals theory obtained $\alpha_{\rm d}=3.31$
for the d, f and g states.  
We adopt $\rho=2.25$ the cut--off parameter. This  yields accurate 
binding energies for the $n$=3,4,5,6,7 and 8 configurations of \ion{Ca}{ii}.

\begin{table}          
\caption[ ]{Term energies for \ion{Ca}{ii} (in Ryd). The table shows results computed without
 PI (w/o PI), with PI and 
 experimental energies from NIST V.3.1.0
}
\begin{flushleft}
\begin{tabular}{lllll}
            \hline\noalign{\smallskip}
 & TERM& w/o PI& PI& NIST\\
            \hline\noalign{\smallskip}

            \hline\noalign{\smallskip}
 1&   4s \ \  $^2$S&0.000000&0.000000&0.000000\\
 2&   3d \ \  $^2$D&0.147449&0.124596&0.124721\\  
 3&   4p \ \  $^2$P$^{\rm o}$&0.219659&0.232385&0.230916\\  
 4&   5s \ \  $^2$S&0.454729&0.478864&0.475380\\
 5&   4d \ \  $^2$D&0.500830&0.524648&0.518062\\   
 6&   5p \ \  $^2$P$^{\rm o}$&0.529136&0.555242&0.552092\\
 7&   4f \ \  $^2$F$^{\rm o}$&0.593486&0.623006&0.620180\\  
 8&   6s \ \  $^2$S&0.618814&0.647386&0.644061\\
 9&   5d \ \  $^2$D&0.639097&0.667632&0.662741\\   
10&   6p \ \  $^2$P$^{\rm o}$&0.652978&0.682121&0.678980\\
11&   5f \ \  $^2$F$^{\rm o}$&0.683576&0.713962&0.711101\\   
12&   5g \ \  $^2$G&0.683935&0.715180&0.712289\\   
13&   7s \ \  $^2$S&0.697115&0.727147&0.723986\\
14&   6d \ \  $^2$D&0.707827&0.737833&0.733791\\  
15&   6f \ \  $^2$F$^{\rm o}$&0.732574&0.763404&0.760526\\  
16&   6g \ \  $^2$G&0.732824&0.764166&0.761272\\  
17&   8s \ \  $^2$S&0.740613&0.771273& 0.768206\\
18&   7d \ \  $^2$D&0.746957&0.777594&0.773987\\  
19&   7f \ \  $^2$F$^{\rm o}$&0.762129&0.793202&0.790315\\  
20&   7g \ \  $^2$G&0.762303&0.793703& 0.790807\\
21&   8d \ \  $^2$D&0.771345&0.802302& 0.798935\\
22&   8f \ \  $^2$F$^{\rm o}$&0.781312&0.812527& 0.809636\\
23&   8g \ \  $^2$G&0.781436&0.812872&0.809975\\
\noalign{\smallskip}
            \hline
         \end{tabular}
\end{flushleft}
\end{table}

The expansion considered here for the \ion{Ca}{ii} system includes 23 LS terms. 
Table 2 presents the complete list
of states included as well as a comparison between the calculated 
and observed target term energies, averaged over fine structure. Here,
we show the energies without polarization interaction (w/o PI)  and those 
with polarization interaction (PI). 
It can be seen that the contribution of PI can reach up to
 15\%, especially for the lower energy terms.   

\begin{figure}         
  \resizebox{\hsize}{!}{ \includegraphics{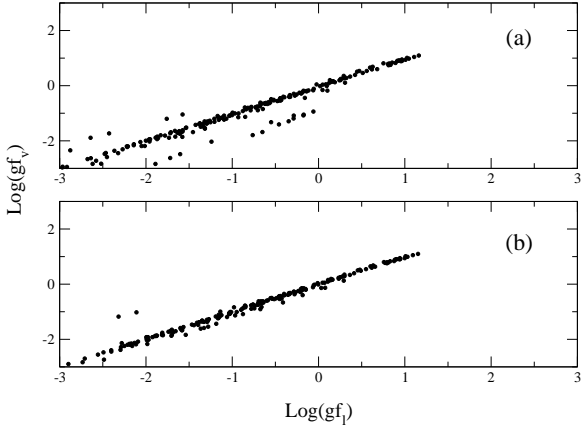}}\caption{ log~$gf_V$ plotted against log~$gf_L$ for transitions between energy levels. 
Panels (a) and (b) show the results computed without polarization interaction
and with polarization interaction respectively}
    \label{EH0}
\end{figure}

In the calculation of radiative rates, 
fine tuning of eigenstates is performed with term energy corrections (TEC), where the improved
relativistic wave function, $\psi_i^R$, is obtained in terms of
the non-relativistic functions
\begin{equation}
\psi_i^R = \psi_i^{LS} + \sum_{j\neq i}\psi_j^{LS}\times \frac{\langle \psi_j^{LS} \mid H_{bp} \mid \psi_i^{LS} \rangle}{E_i^{LS}
-E_j^{LS}},
 \label{dos}
\end{equation}
with the $LS$ energy differences
$E_i^{LS}-E_j^{LS}$ adjusted to fit weighted averaged energies of the experimental multiplets \citep{1977MNRAS.181..527Z}.

In our best target representation, which accounts for the interaction between the valence 
electron and the core, the theoretical energies for 
all the  23 terms are typically within
2\% of the experimental values before any further empirical correction. 
After TEC, the agreement with experimental energies is better than 1\%. 

For dipole-allowed transitions, spontaneous decay rates are given by
\begin{equation}
A_{ij}^{E1}=2.6774\times  10^9(E_i-E_j)^3\frac{1}{g_{i}}S^{E1}_{ij} \ \ (s^{-1}), \label{tres}
\end{equation}
while for forbidden transitions we consider electric quadrupole (E2)
and magnetic dipole (M1) transition rates given by
\begin{equation}
A_{ij}^{E2}=2.6733\times  10^3(E_i-E_j)^5\frac{1}{g_{i}}S^{E2}_{ij} \ \ (s^{-1})
 \label{cuatro}
\end{equation}
and
\begin{equation}
A_{ij}^{M1}=3.5644\times  10^4(E_i-E_j)^3\frac{1}{g_{i}}S^{M1}_{ij}  \ \ (s^{-1}). \label{cinco}
\end{equation}
Here, $g_i$ is the statistical weight of the upper initial level 
$i$, $S_{ij}$ is the line strength and $E$ is the energy
in Rydbergs.

Eqns.(\ref{tres},\ref{cuatro} and \ref{cinco}) show that the transition rates are sensitive to the 
accuracy of the energy levels, particularly for forbidden transitions among 
nearby levels. Thus, we perform further adjustments to the transitions
rates by correcting our best calculated energies to experimental values.
 
In Fig. \ref{EH0} we plot the $gf$-values for dipole allowed transitions
among fine structure levels computed
in the length gauge vs. those in the velocity gauge. We present  the 
$gf$-values without PI 
 (a) and with PI (b). The overall agreement
between the two gauges is around 5\% for $\log(gf)$-values greater than
$-3$ when accounting for PI and greater than 15\% without PI.
 This is a good indicator of the quality of the dipole allowed radiative
data. 

In Table \ref{ft} we present an extensive comparison between the present
results and previous lifetimes for the metastable levels
${\rm 3d~^2D_{5/2}}$ and ${\rm 3d~^2D_{3/2}}$. These levels are of particular
astrophysical interest because they yield the prominent spectral lines
$\lambda \lambda$ 7293,~7326 \AA. Our results including PI
and TEC are in excellent agreement with  experimental values, while
the results that neglect PI are $\sim$10\% too low.

\begin{table}[h]
\caption{Lifetimes of the metastable ${\rm 3d~^2D}$ levels of \ion{Ca}{ii}}  
\begin{flushleft}
\begin{tabular}{lllll}
            \hline\noalign{\smallskip}
Level&\multicolumn{2}{c}{Present}&Other&Experiment($\tau$(s))\\
&w/o PI&PI & & \\
            \hline\noalign{\smallskip}
${\rm 3d~^2D_{3/2}}$&0.926&1.107&1.081$^1$&1.176$\pm$0.011$^5$\\
              &     &     &1.16$^2$&\\
              &     &     &1.27$^3$&\\
              &     &     &0.98$^4$&\\
              &     &     &1.196$^5$&\\
${\rm 3d~^2D_{5/2}}$&0.901&1.08&1.058$^1$&1.168$\pm$0.009$^5$\\
              &     &     &1.14$^2$&1.152$\pm$0.020$^7$\\
              &     &     &1.24$^3$&1.100$\pm$0.018$^8$\\
              &     &     &0.95$^4$&1.054$\pm$0.061$^9$\\
              &     &     &1.165$^5$&1.149$\pm$0.014$^{10}$\\
              &     &     &1.045$^6$&1.064$\pm$0.017$^{11}$\\
              &     &     &1.14$^{11}$& \\
\noalign{\smallskip}
            \hline
         \end{tabular}
\end{flushleft}
$^1$\cite{1990AA...229..248Z},  
$^2$\cite{1992PhRvA..46.3704V}, 
$^3$\cite{1991PhRvA..44.1531G}, 
$^4$\cite{1988PhRvA..38.3992A}, 
$^5$\cite{2005PhRvA..71c2504K}, 
$^6$\cite{1995PhRvA..51.1723L}, 
$^7$\cite{2004EPJD...29..163K}, 
$^8$\cite{1999EPJD....7..461B}, 
$^9$\cite{1994ZPhyD..29..159A}, 
$^{10}$\cite{2004PhRvA..69c2503S}, 
$^{11}$\cite{1996EL.....33..595G}\\ 
\label{ft}
\end{table}

In Table \ref{lt1} we compare the calculated lifetimes for short-lived
levels of \ion{Ca}{ii} from the present
calculations  with other theoretical
and experimental values.
 For the lower levels ($4p~^2P^{\rm o}_{1/2}$ and $4p~^2P^{\rm o}_{3/2}$) the effect
of polarization interaction is $\sim$20\%. Overall, the differences
between the results of our best model and experimental values are less than
5\%, except for the level $5d~^2D_{5/2}$. For this level, our result agrees
with previous independent calculations but
is about 40\% below the experimental values of \cite{1970JOSA...60.1199A}.
A new measurement of this lifetime would be very important.

\begin{table*}
\caption[ ]{\ion{Ca}{ii} lifetimes (in ns). The second 
 column gives the results with neither PI (w/o PI) nor 
 TECs (w/o TECs), the third column 
gives the results without PI (w/o PI) but with TEC,
the fourth columns gives results with PI but no TEC, and the fifth column 
 shows the results with both PI and TEC.
Theoretical (Other) and experimental (Experiment) values from other authors are given in
the sixth and seventh columns, respectively.}  
\begin{flushleft}
\begin{tabular}{lllllll}
            \hline\noalign{\smallskip}
Level & \multicolumn{4}{c}{Present} &Other & Experiment\\
&\multicolumn{2}{c}{w/o PI}&\multicolumn{2}{c}{PI}&&\\
&w/o TEC & TEC &w/o TEC&TEC & & \\ 
            \hline\noalign{\smallskip}
${\rm 4p~^2P^o_{1/2}}$&6.978&5.734&6.697&6.837 &     6.44$^a$& 7.07$\pm 0.07 ^b$\\
          &&&      &&      6.87$^b$ & 7.5$\pm$0.5$^g$\\
          &&&      &&      6.39$^c$ & 6.62$\pm$0.35$^h$\\ 
          &&&      &&      6.94$^d$ & 6.95$\pm$0.18$^i$\\
          &&&      &&      7.045$^e$& 7.098$\pm$0.020$^{m}$\\
          &&&      &&      7.047$^{l}$& \\
${\rm 4p~^2P^o_{3/2}}$&6.797&5.577&6.508&6.649 &     6.28$^a$& 6.87$\pm$0.06$^b$\\
          &&&      &&      6.24$^b$ & 7.4$\pm$0.6$^g$\\
          &&&      &&      6.21$^c$ & 6.68$\pm$0.35$^h$\\
          &&&      &&      6.75$^d$ & 6.72$\pm$0.20$^{j}$\\
          &&&      &&     6.852$^e$ & 6.61$\pm$0.30$^{k}$\\
          &&&      &&     6.833$^{l}$ & 6.87$\pm$0.18$^i$\\
          &&&      &&               &6.924$\pm$0.019$^{m}$ \\
${\rm 4d~^2D_{3/2}}$&2.963&2.779&2.781&2.934& 2.868$^e$ &\\
           
${\rm 4d~^2D_{5/2}}$&2.981&2.799&2.800&2.952& 2.886$^e$ &\\

${\rm 4f~^2F^o_{5/2}}$&3.625&2.656&3.487&3.451& 3.895$^e$ &\\

${\rm 4f~^2F^o_{7/2}}$&3.625&2.654&3.486&3.448& 3.897$^e$ &\\

${\rm 5s~^2S_{1/2}}$&4.310&3.833&3.886&3.982& 4.13$^a$&4.3$\pm$0.4$^g$\\
          &&&      &&        4.153$^e$ & \\
          &&&      &&       3.85$^f$ &\\
${\rm 5p~^2P^o_{1/2}}$&40.195&34.903&36.541&35.174& 33.78$^a$ &\\
          &&&      &&       36.200$^e$ & \\

${\rm 5p~^2P^o_{3/2}}$&38.987&33.755&35.783&34.401& 33.92$^a$ &\\
          &&&      &&       35.349$^e$ &\\

${\rm 5d~^2D_{3/2}}$&5.960&5.625&5.895&6.102& 6.148$^e$ &      \\

${\rm 5d~^2D_{5/2}}$&6.002&5.669&5.944&6.154& 6.199$^e$ & 4.3$\pm$0.2$^g$\\

${\rm 6s~^2S_{1/2}}$&7.007&6.390&6.391&6.457&  6.90$^a$ & \\
          &&&       &&        6.766$^e$ &\\
          &&&       &&        6.51$^f$ &\\

${\rm 6p~^2P^o_{1/2}}$&114.06&98.569&93.141&90.173& 58.95 $^a$ \\
          &&&       &&       100.254$^e$\\

${\rm 6p~^2P^o_{3/2}}$&111.23&95.788&92.262&89.129& 43.06 $^a$  \\
          &&&       &&       99.675$^e$\\
\noalign{\smallskip}
            \hline
         \end{tabular}
\end{flushleft}
$^a$\cite{1978JPhB...11.2975H},
$^b$\cite{1988PhRvA..38.4887G},
$^c$\cite{1969atp..book.....W},
$^d$\cite{1991PhRvA..44.1531G},
$^e$\cite{1989PhRvA..39.4880T},
$^f$\cite{1972ApJ...177..567B},
$^g$\cite{1970JOSA...60.1199A},
$^h$\cite{1985PhLA..110..383A},
$^i$\cite{1988NIM31305},
$^{j}$\cite{1966PhRv..145...26S},
$^{k}$\cite{1976PhRvA..14.1735R},
$^{l}$\cite{1995PhRvA..51.1723L},
$^{m}$\cite{1993PhRvL..70.3213J}\\
\label{lt1}
\end{table*}

In Table \ref{fg} we present a comparison between calculated and experimental oscillator
strengths in absorption. For the calculated oscillator strengths we choose the most complete and representative
set of values as well as the most cited. We compare with previous calculations by \cite{1992PhRvA..46.3704V} where
they use  a multiconfiguration Hartree-Fock method with core polarization included
 variationally (SECP) 
using a model potential of the form described first by \cite{baylis77} and then reviewed by \cite{hibbert89}.
We also compare with 
\cite{1989PhRvA..39.4880T} who used a Hartree-Slater (HS) core potential
 with semiempirical corrections, \cite{1991PhRvA..44.1531G} who used relativistic many-body theory and \cite{1993PhyS...48..533B} 
 using an orthogonal Breit-Pauli MCHF method. We also
compare with experimental values of \cite{1967PhRv..157...24G} who used the Hanle-effect technique with optical excitation from the ion ground state. The inclusion of PI can affect the oscillator 
strengths by up to 12\%. One finds very
 good agreement between the present calculation, previous calculations and experimental values.

\begin{table*}
\caption[ ]{Oscillator strengths  for \ion{Ca}{ii}. The oscillator strengths from previous calculations of 
\cite{1992PhRvA..46.3704V}, \cite{1989PhRvA..39.4880T}, \cite{1991PhRvA..44.1531G} and \cite{1993PhyS...48..533B}, and experimental data from
\cite{1967PhRv..157...24G} are also given. 
 }

\begin{flushleft}
\begin{tabular}{llllllll}
            \hline\noalign{\smallskip}
Transition &\multicolumn{2}{c}{Present}&Vaeck&Theodosiou&Guet&Brage& Experiments\\
&w/o PI&PI&\\
            \hline\noalign{\smallskip}

     ${\rm 4s~S_{1/2}-4p~P^o_{1/2}}$    &0.364  &0.323   &0.318 &0.316 &0.320 & 0.321\\
     ${\rm 4s~^2S_{1/2}-4p~^2P^o_{3/2}}$&0.734  &0.652   &0.641 &0.637 &0.645 &0.649 &0.66$\pm$0.02 \\
     ${\rm 3d~^2D_{3/2}-4p~^2P^o_{1/2}}$&0.0478 &0.0537  &0.0547&0.0473&0.0494&0.0524& \\
     ${\rm 3d~^2D_{3/2}-4p~^2P^o_{3/2}}$&0.0098 &0.0110  &0.0112&0.0096&0.0101&0.0107&0.0088$\pm$0.001 \\
     ${\rm 3d~^2D_{5/2}-4p~^2P^o_{3/2}}$&0.0584 &0.0656  &0.0666&0.0574&0.0601&0.0637&0.053$\pm$0.006 \\
\noalign{\smallskip}
            \hline
         \end{tabular}
\end{flushleft}
\label{fg}
\end{table*}

In Table \ref{E2} we present  our electric quadrupole and magnetic dipole     
transitions probabilities 
computed including PI and TEC. We compare these 
with previous calculations of \cite{1990AA...229..248Z}, who used a two-step minimization procedure and semi-empirical term energy corrections 
using the computer program SUPERSTRUCTURE, \cite{1988PhRvA..38.3992A} who used the multiconfiguration Dirac-Fock 
formalism and \cite{1992PhRvA..46.3704V}. Our results are in good agreement (within $\sim$2\%) with \cite{1990AA...229..248Z}
and within $\sim$13\% and $\sim$8\% with respect to \cite{1988PhRvA..38.3992A} and \cite{1992PhRvA..46.3704V}. 

\begin{table*}\caption[ ]{Electric quadrupole (E2) and magnetic dipole (M1) transitions probabilities of [\ion{Ca}{ii}]. We compare our results with previous calculated values of \cite{1990AA...229..248Z}, \cite{1988PhRvA..38.3992A} and \cite{1992PhRvA..46.3704V}. } \begin{flushleft}
\begin{tabular}{llllll}
            \hline
A($s^{-1}$)  &  Type & Present & Zeippen& Ali and Kim&Vaeck\\
            \hline\noalign{\smallskip}
${\rm 3d~^2D_{3/2} - 4s~^2S_{1/2}}$& E2  &0.905&0.925&1.02&0.84 \\${\rm 3d~^2D_{5/2} - 4s~^2S_{1/2}}$& E2  &0.928&0.945&1.05&0.86 \\
${\rm 3d~^2D_{5/2} - 3d~^2D_{3/2}}$& M1  &2.41$\times 10^{-6}$&2.45$\times 10^{-6}$&2.45$\times 10^{-6}$& \\
            \hline\noalign{\smallskip}         \end{tabular}
\end{flushleft}\label{E2}
\end{table*}

\subsection{Scattering calculations}
In the close coupling (CC) approximation the total wave function of the 
electron-ion system is represented as
\begin{equation}
\psi(E;LS\pi)=A\sum_i \chi_i\theta_i + \sum_jc_j\Phi_j
 \label{seis}
\end{equation}
where $\chi_i$ is the target ion wave function in a specific state 
$LS_i$, $\theta_i$ is the wave function of the free
electron, $\Phi_j$ are short range correlation functions for the 
bound (e+ion) system, and $A$ is the antisymetrization operator.

The variational procedure gives rise to a set of coupled integro-differential 
equations that are solved with the R-matrix technique 
\citep{1971JPhB....4..153B, 1978CoPhC..14..367B, 1995CoPhC..92..290B} within 
a box of radius $r\le a$. In the asymptotic region, $r>a$, exchange between the outer electron and 
the target ion can be neglected and if all long-range potentials beyond 
Coulombic are also neglected, the reactance $K$-matrix and the scattering 
$S$-matrix are obtained by matching at the boundary the inner-radial 
functions to linear combinations of the outer-region Coulomb solutions.
Later, the contributions of long-range potentials to the collision strengths
are included perturbatively \citep[]{1998JPhB...31.3713G}.

We use the $LS$-coupling R-matrix method that includes mass-velocity and Darwin operators 
and the polarization model  potential.
Note that the scattering calculations include one-body relativistic operators
while the atomic structure calculations using AUTOSTRUCTURE include two-body operators of the
 Breit-Pauli Hamiltonian namely (two-body) fine-structure operators and (two-body) non-fine-structure operators.

The $S$-matrix elements determine the collision strength for a transition from an initial target state $i$ to a final target state $f$,
\begin{equation}
\Omega_{if}=\frac{1}{2}\sum w|S_{if}-\delta_{if}|,
 \label{siete}\end{equation}
where $w=(2L+1)(2S+1)$ or $(2J+1)$ depending on the coupling scheme, and the summation runs over the partial waves and
channels coupling the initial and final states of interest.

In order to derive fine-structure results out of the $LS$-coupling calculation
we employ the intermediate-coupling frame transformation (ICFT)
method of \cite{1998JPhB...31.3713G}.
The ICFT method uses the multi-channel quantum defect theory (MQDT) to generate the LS-coupled unphysical $K$-matrices. 
In this approach one treats all scattering channels 
as open and calculates the term-coupling coefficients (TCCs) to transform the unphysical $K$-matrices 
 to full intermediate coupling. Finally, we can generate the physical $K$-matrices on a fine energy mesh.
Because all channels are treated as open 
this method eliminates the problems associated with the transformation of 
the physical $S$-matrices with closed channels 
and consequently yields accurate results for  both background and 
resonances of the collision strengths at all energies.

The computations were carried out with the RMATRX package of codes
\citep{1995CoPhC..92..290B} and also include the dipole polarization potential for the interaction of the valence electron
and the core. The set of (N+1)-electron wave functions to the right of the CC expansion in
Eq.~(\ref{seis}) includes all the configurations that result from adding
an additional electron to the target configurations.  

\begin{figure}
  \resizebox{\hsize}{!}{\includegraphics{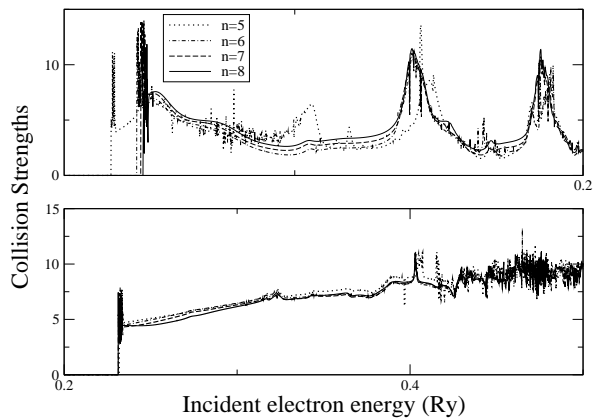}}
\caption{Comparison between collision strengths for 
 the \ion{Ca}{ii} ion with different CC expansions for
the ${\rm 3d~^2D_{3/2}-4s~^2S_{1/2}}$ (upper panel) and ${\rm 4p~^2P^o_{1/2}-4s~^2S_{1/2}}$
(lower panel) transitions. $n$ represents the principal quantum number for the valence electron.}
    \label{cc3}
\end{figure}

In Fig.~\ref{cc3} we compare the collision strengths
 for different target expansions. 
For the various close-coupling  expansions we include all the terms 
that arise from the $n$=3, 4, 5, 6, 7 and 8
 configurations.   
It is interesting to note that by increasing the number of terms in the close-coupling expansion 
the resonances tend to become better organized and blended, leading
to regular broad series of structures. 
This is because of the increasing number of channels for 
decay of autoionizing levels. 
We adopt the expansion up to n=8 for all further calculations.

Partial wave contributions to the summation in Eqn.~3 were included from 162 $SL\pi$
total symmetries with angular momentum $L = 0 - 40$, total multiplicities
$(2S+1) = 1$ and 3, and parities even and odd.
As for the number of continuum orbitals per angular momentum 
we found necessary to include up to $40$.
The  collision strengths were ``topped up" with estimates of the
contributions of higher partial waves for the optically allowed transitions
based on the Coulomb-Bethe
approximation \cite{1974JPhB....7L.364B} and for non-allowed transitions we approximate
the top-up with a geometric series.
It was verified that by explicitly including partial waves up to $L = 40$
such ``top ups" amount to less than 10\% of the total collision strengths
for all transitions with the exception of the transitions 
$n$g-$n$f (for $n>$6) for which the top up is greater than $20 \%$.
This is because of the very small energy difference between these terms.

\begin{figure}
  \resizebox{\hsize}{!}{\includegraphics{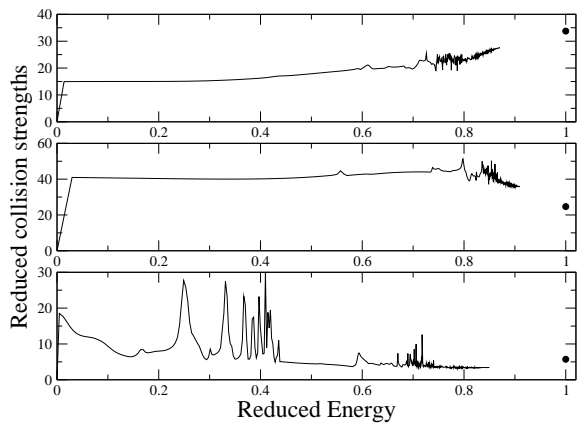}}
    \caption{Reduced collision strengths for the ${\rm 4s~^2S~-~4p~^2P^o}$ (upper panel),
${\rm 3d~^2D~-~4p~^2P^o}$ (middle panel) and ${\rm 4s~^2S~-~3d~^2D}$ (lower panel) transitions.}
    \label{xb1}
\end{figure}

 In order to check for convergence of the partial wave expansion 
one needs to study the high energy behavior for the collision strengths ($\Omega$). In Fig.~\ref{xb1}
we plot reduced collision strengths ($\Omega_{r}$) as function of reduced energy ($E_r$) following the
procedure describe by \cite{1992AA...254..436B}. This approach allows us to visualize the complete range of
energies mapped onto the interval [0,1]. The last point in these graphs represents the infinite energy limit of
the collision strengths ($E \rightarrow \infty$). For this plot we
adopt a scaling parameter $C=1.5$.
These plots show good progression of the collision strengths toward
the high energy limit, which gives confidence on the
consistency and quality of the data.

\begin{figure*}
 \centering
  \includegraphics[width=17cm]{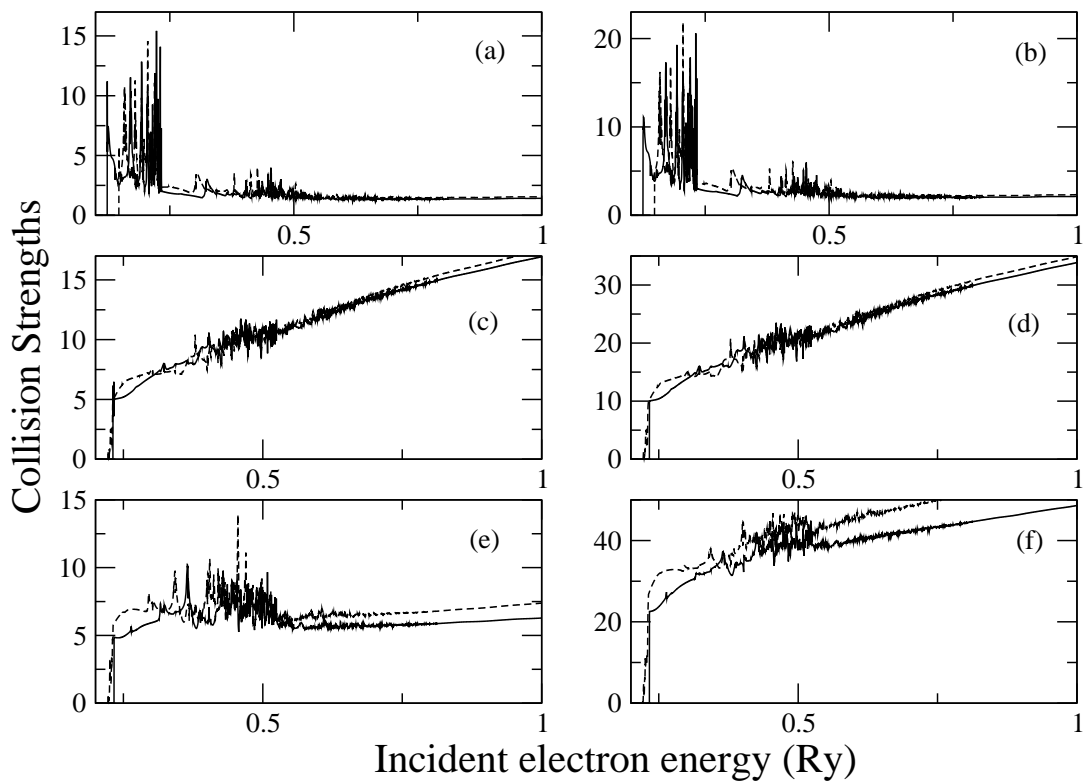}
\caption{Comparison between collision strengths for  the \ion{Ca}{ii} ion. The
solid and dashed lines depict
the results with and without model potential interaction respectively.
The transitions are: (a) ${\rm 3d~^2D_{3/2}-4s~^2S_{1/2}}$;
(b) ${\rm 3d~^2D_{5/2}-4s~^2S_{1/2}}$; (c) ${\rm 4p~^2P^o_{1/2}-4s~^2S_{1/2}}$; (d) ${\rm 4p~^2P^o_{3/2}-4s~^2S_{1/2}}$;
 (e) ${\rm 3d~^2D_{3/2}-4p~^2P^o_{3/2}}$ and (f) ${\rm 3d~^2D_{5/2}-4p~^2P^o_{3/2}}$.}
\label{c1} 
\end{figure*}

Fig.~\ref{c1} shows the collision strength computed with and without 
polarization interaction for a sample of transitions between levels of the 
lowest four multiplets. One can see that the core polarization has an important 
effect on the collision strengths. We verified through various calculations that 
such effects come almost entirely from the change in the radial wave functions 
of the target, while the effects of the dipole polarization operator in the 
scattering matrix is nearly negligible.
The same conclusion was derived by \cite{1988PhRvA..38.3339M}. 

The collision strengths were calculated at $22000$ energy points from 0 to 4 Ry,
with a resolution of $10^{-5}$~Ry in the region with resonances and
$5\times10^{-3}$~Ry at higher energies.
This number of points was found
sufficient to resolve most resonance structures for accurate calculations of effective collision strengths for temperatures above $\sim$5000~K.

A dimensionless thermally-averaged effective collision strength results from 
integrating the collision
 strength over a Maxwellian distribution of electron velocities
\begin{equation}
\Upsilon_{if} = \int_0^\infty \Omega_{if} \exp{(-\epsilon_f/kT)}
$d$(\epsilon_f/kT),
 \label{ocho}
\end{equation}
where $\epsilon_f$ is the kinetic energy of the outgoing electron, $T$
the electron temperature in Kelvin and $k=6.339\times 10^{-6}$
Ry/K is the Boltzmann constant.

In Table \ref{coliefe} we compare the present effective collision strengths for the 4s$\to$ 4p 
 transition of \ion{Ca}{ii}. We show the values obtained by \cite{1977ApL....19...11O}  
 where they used experimental results of \cite{1973PhRvA...8.2304T}, as reduced to collision strengths by \cite{seaton1975}.
We also present the values from \cite{1995A&A...300..627B} as we use their 
five point cubic spline fitting parameters to tabulate the $\Upsilon (T)$ using the procedure described in \cite{1992AA...254..436B}.
 We compare them with our present calculation without PI in the target representation and in the scattering
 calculations  (w/o PI); without PI in the R-matrix calculations only (PI-S) and with PI.
Our best results agree within 11\% of \cite{1977ApL....19...11O}  values.
 On the other hand, the results of \cite{1995A&A...300..627B} obtained with the
non-exchange distorted wave approximation appear overestimated by
$\sim$50\%. In this calculation, the \ion{Ca}{II} target is represented in the 
frozen core approximation neglecting polarization. 
As we pointed out before
the main contribution of the dipole polarization potential is in the
representation of the target ion. 
Further, our effective collision strengths when neglecting
polarization are overestimated by $30\%$, in closer agreement with 
\cite{1995A&A...300..627B}.

\begin{table}
\caption[ ]{Effective collision strengths 
for the 4s~$^2$S ~-~4p~$^2$P$^{\rm o}$ transition
\ion{Ca}{ii}. The first three lines correspond to results of the present
calculations for the cases of no PI in either
the target orbitals of scattering calculation (w/o PI), PI 
in the scattering calculation only (PI-S), and 
PI in the target and scattering calculations (PI). These
results are compared with values deduced from experimental cross sections
\citep{1977ApL....19...11O} (OW) and the most recent theoretical calculations 
\citep{1995A&A...300..627B} (BCT). } 
\begin{flushleft}
\begin{tabular}{lllll}
            \hline\noalign{\smallskip}
\multicolumn{4}{c}{$T~(k)$}&\\
5000&10000&15000&20000&\\
            \hline\noalign{\smallskip}
19.03&20.82&22.45&24.03&w/o PI \\
16.96&19.34&21.35&23.13&PI-S\\
17.07&19.44&21.45&23.23&PI\\
15.6&17.5&19.2&20.8&OW\\
24.87&27.50&29.88&32.04&BCT\\
\noalign{\smallskip}
            \hline
         \end{tabular}
\end{flushleft}
\label{coliefe}
\end{table}

\section{Conclusions}

We have computed radiative data, collision strengths and effective collision strengths 
for transitions among 41 levels from the n=3, 4, 5, 6, 7 and 8 configurations of \ion{Ca}{ii}.
The radiative data were calculated using the Thomas-Fermi-Dirac central potential
with a model core potential that account for dipole 
polarization interaction of the valence electron with the core. We also present an extensive comparison between 
our results and the most recent experiments and calculations for the lifetimes of \ion{Ca}{ii}. 
 
Effective collision strengths are available 
for various temperatures that expand from 3000~K to 38000~K.
The whole set of data reported here including energy levels, infinite energy limit Born collision strengths,
 transition probabilities and effective collision strengths can be obtained in
electronic form at the CDS via anonymous ftp to cdsarc.u-strasbg.fr (130.79.128.5), via http://cdsweb.u-strasbg.fr/cgi-bin/qcat?J/A+A/
 or by request to the authors.


\bibliographystyle{aa}
\bibliography{ca2}

\begin{thebibliography}{60}
\expandafter\ifx\csname natexlab\endcsname\relax\def\natexlab#1{#1}\fi

\bibitem[{{Ali} \& {Kim}(1988)}]{1988PhRvA..38.3992A}
{Ali}, M.~A. \& {Kim}, Y.-K. 1988, \pra, 38, 3992

\bibitem[{{Andersen} {et~al.}(1970){Andersen}, {Desesquelles}, {Jessen}, \&
  {Sorensen}}]{1970JOSA...60.1199A}
{Andersen}, T., {Desesquelles}, J., {Jessen}, K.~A., \& {Sorensen}, G. 1970, J.
  Quant. Spectrosc. Radiat. Transfer., 10, 1143

\bibitem[{{Ansbacher} {et~al.}(1985){Ansbacher}, {Inamdar}, \&
  {Pinnington}}]{1985PhLA..110..383A}
{Ansbacher}, W., {Inamdar}, A.~S., \& {Pinnington}, E.~H. 1985, Physics Letters
  A, 110, 383

\bibitem[{{Arbes} {et~al.}(1994){Arbes}, {Benzing}, {Gudjons}, {Kurth}, \&
  {Werth}}]{1994ZPhyD..29..159A}
{Arbes}, F., {Benzing}, M., {Gudjons}, T., {Kurth}, F., \& {Werth}, G. 1994,
  Zeitschrift fur Physik D Atoms Molecules Clusters, 29, 159

\bibitem[{{Badnell}(1986)}]{1986JPhB...19.3827B}
{Badnell}, N.~R. 1986, J. Phys. B: At. Mol. Opt. Phys, 19, 3827

\bibitem[{{Badnell}(1997)}]{1997JPhB...30....1B}
{Badnell}, N.~R. 1997, J. Phys. B: At. Mol. Opt. Phys, 30, 1

\bibitem[{{Bautista}(2001)}]{bau01}
{Bautista}, M.~A. 2001, \aap, 365, 268

\bibitem[{{Bautista}(2004)}]{bau04}
{Bautista}, M.~A. 2004, \aap, 420, 763

\bibitem[{{Bautista} \& {Pradhan}(1998)}]{1998RMxAC...7..163B}
{Bautista}, M.~A. \& {Pradhan}, A.~K. 1998, in Revista Mexicana de Astronomia y
  Astrofisica Conference Series, ed. R.~J. {Dufour} \& S.~{Torres-Peimbert},
  163

\bibitem[{{Baylis}(1977)}]{baylis77}
{Baylis}, W.~E. 1977, J. Phys. B: At. Mol. Opt. Phys, 10, L583

\bibitem[{{Berrington} {et~al.}(1978){Berrington}, {Burke}, {Le Dourneuf},
  {Robb}, {Taylor}, \& {Ky Lan}}]{1978CoPhC..14..367B}
{Berrington}, K.~A., {Burke}, P.~G., {Le Dourneuf}, M., {et~al.} 1978, Computer
  Physics Communications, 14, 367

\bibitem[{{Berrington} {et~al.}(1995){Berrington}, {Eissner}, \&
  {Norrington}}]{1995CoPhC..92..290B}
{Berrington}, K.~A., {Eissner}, W.~B., \& {Norrington}, P.~H. 1995, Computer
  Physics Communications, 92, 290

\bibitem[{{Black} {et~al.}(1972){Black}, {Weisheit}, \&
  {Laviana}}]{1972ApJ...177..567B}
{Black}, J.~H., {Weisheit}, J.~C., \& {Laviana}, E. 1972, \apj, 177, 567

\bibitem[{{Block} {et~al.}(1999){Block}, {Rehm}, {Seibert}, \&
  {Werth}}]{1999EPJD....7..461B}
{Block}, M., {Rehm}, O., {Seibert}, P., \& {Werth}, G. 1999, European Physical
  Journal D, 7, 461

\bibitem[{{Brage} {et~al.}(1993){Brage}, {Froese Fischer}, {Vaeck},
  {Godefroid}, \& {Hibbert}}]{1993PhyS...48..533B}
{Brage}, T., {Froese Fischer}, C., {Vaeck}, N., {Godefroid}, M., \& {Hibbert},
  A. 1993, \physscr, 48, 533

\bibitem[{{Burgess}(1974)}]{1974JPhB....7L.364B}
{Burgess}, A. 1974, J. Phys. B: At. Mol. Opt. Phys, 7, L364

\bibitem[{{Burgess} {et~al.}(1995){Burgess}, {Chidichimo}, \&
  {Tully}}]{1995A&A...300..627B}
{Burgess}, A., {Chidichimo}, M.~C., \& {Tully}, J.~A. 1995, \aap, 300, 627

\bibitem[{{Burgess} \& {Tully}(1992)}]{1992AA...254..436B}
{Burgess}, A. \& {Tully}, J.~A. 1992, \aap, 254, 436

\bibitem[{{Burke} {et~al.}(1971){Burke}, {Hibbert}, \&
  {Robb}}]{1971JPhB....4..153B}
{Burke}, P.~G., {Hibbert}, A., \& {Robb}, W.~D. 1971, J. Phys. B: At. Mol. Opt.
  Phys, 4, 153

\bibitem[{{Chidichimo}(1981)}]{1981JPhB...14.4149C}
{Chidichimo}, M.~C. 1981, J. Phys. B: At. Mol. Opt. Phys, 14, 4149

\bibitem[{{Eissner} {et~al.}(1974){Eissner}, {Jones}, \&
  {Nussbaumer}}]{1974CoPhC...8..270E}
{Eissner}, W., {Jones}, M., \& {Nussbaumer}, H. 1974, Computer Physics
  Communications, 8, 270

\bibitem[{{Eissner} \& {Nussbaumer}(1969)}]{1969JPhB....2.1028E}
{Eissner}, W. \& {Nussbaumer}, H. 1969, J. Phys. B: At. Mol. Opt. Phys, 2, 1028

\bibitem[{{Ferland} \& {Persson}(1989)}]{1989ApJ...347..656F}
{Ferland}, G.~J. \& {Persson}, S.~E. 1989, \apj, 347, 656

\bibitem[{{Gallagher}(1967)}]{1967PhRv..157...24G}
{Gallagher}, A. 1967, Physical Review, 157, 24

\bibitem[{{Gosselin} {et~al.}(1988{\natexlab{a}}){Gosselin}, {Pinnington}, \&
  {Ansbacher}}]{1988PhRvA..38.4887G}
{Gosselin}, R.~N., {Pinnington}, E.~H., \& {Ansbacher}, W. 1988{\natexlab{a}},
  \pra, 38, 4887

\bibitem[{{Gosselin} {et~al.}(1988{\natexlab{b}}){Gosselin}, {Pinnington}, \&
  {Ansbacher}}]{1988NIM31305}
{Gosselin}, R.~N., {Pinnington}, E.~H., \& {Ansbacher}, W. 1988{\natexlab{b}},
  Nuclear. Instrum. Methods, 31, 305

\bibitem[{{Griffin} {et~al.}(1998){Griffin}, {Badnell}, \&
  {Pindzola}}]{1998JPhB...31.3713G}
{Griffin}, D.~C., {Badnell}, N.~R., \& {Pindzola}, M.~S. 1998, J. Phys. B: At.
  Mol. Opt. Phys, 31, 3713

\bibitem[{{Gudjons} {et~al.}(1996){Gudjons}, {Hilbert}, {Seibert}, \&
  {Werth}}]{1996EL.....33..595G}
{Gudjons}, T., {Hilbert}, B., {Seibert}, P., \& {Werth}, G. 1996, Europhysics
  Letters, 33, 595

\bibitem[{{Guet} \& {Johnson}(1991)}]{1991PhRvA..44.1531G}
{Guet}, C. \& {Johnson}, W.~R. 1991, \pra, 44, 1531

\bibitem[{{Hafner} \& {Schwarz}(1978)}]{1978JPhB...11.2975H}
{Hafner}, P. \& {Schwarz}, W.~H.~E. 1978, J. Phys. B: At. Mol. Opt. Phys, 11,
  2975

\bibitem[{{Hibbert}(1989)}]{hibbert89}
{Hibbert}, A. 1989, \physscr, 39, 574

\bibitem[{{Hummer} {et~al.}(1993){Hummer}, {Berrington}, {Eissner}, {Pradhan},
  {Saraph}, \& {Tully}}]{1993A&A...279..298H}
{Hummer}, D.~G., {Berrington}, K.~A., {Eissner}, W., {et~al.} 1993, \aap, 279,
  298

\bibitem[{{Jin} \& {Church}(1993)}]{1993PhRvL..70.3213J}
{Jin}, J. \& {Church}, D.~A. 1993, Physical Review Letters, 70, 3213

\bibitem[{{Joly}(1989)}]{1989A&A...208...47J}
{Joly}, M. 1989, \aap, 208, 47

\bibitem[{{Jones}(1970)}]{1970JPhB....3.1571J}
{Jones}, M. 1970, J. Phys. B: At. Mol. Opt. Phys, 3, 1571

\bibitem[{{Jones}(1971)}]{1971JPhB....4.1422J}
{Jones}, M. 1971, J. Phys. B: At. Mol. Opt. Phys, 4, 1422

\bibitem[{{Kennedy} {et~al.}(1978){Kennedy}, {Myerscough}, \&
  {McDowell}}]{1978JPhB...11.1303K}
{Kennedy}, J.~V., {Myerscough}, V.~P., \& {McDowell}, M.~R.~C. 1978, J. Phys.
  B: At. Mol. Opt. Phys, 11, 1303

\bibitem[{{Knoop} {et~al.}(2004){Knoop}, {Champenois}, {Hagel}, {Houssin},
  {Lisowski}, {Vedel}, \& {Vedel}}]{2004EPJD...29..163K}
{Knoop}, M., {Champenois}, C., {Hagel}, G., {et~al.} 2004, European Physical
  Journal D, 29, 163

\bibitem[{{Kreuter} {et~al.}(2005){Kreuter}, {Becher}, {Lancaster}, {Mundt},
  {Russo}, {H{\"a}ffner}, {Roos}, {H{\"a}nsel}, {Schmidt-Kaler}, {Blatt}, \&
  {Safronova}}]{2005PhRvA..71c2504K}
{Kreuter}, A., {Becher}, C., {Lancaster}, G.~P., {et~al.} 2005, \pra, 71,
  032504

\bibitem[{{Liaw}(1995)}]{1995PhRvA..51.1723L}
{Liaw}, S.-S. 1995, \pra, 51, 1723

\bibitem[{{Mel{\'e}ndez} \& {Bautista}(2005)}]{2005A&A...436.1123M}
{Mel{\'e}ndez}, M. \& {Bautista}, M.~A. 2005, \aap, 436, 1123

\bibitem[{{Mitroy} {et~al.}(1988){Mitroy}, {Griffin}, {Norcross}, \&
  {Pindzola}}]{1988PhRvA..38.3339M}
{Mitroy}, J., {Griffin}, D.~C., {Norcross}, D.~W., \& {Pindzola}, M.~S. 1988,
  \pra, 38, 3339

\bibitem[{{Norcross} \& {Seaton}(1976)}]{1976JPhB....9.2983N}
{Norcross}, D.~W. \& {Seaton}, M.~J. 1976, J. Phys. B: At. Mol. Opt. Phys, 9,
  2983

\bibitem[{{Osterbrock} \& {Wallace}(1977)}]{1977ApL....19...11O}
{Osterbrock}, D.~E. \& {Wallace}, R.~K. 1977, \aplett, 19, 11

\bibitem[{{Rambow} \& {Schearer}(1976)}]{1976PhRvA..14.1735R}
{Rambow}, F.~H.~K. \& {Schearer}, L.~D. 1976, \pra, 14, 1735

\bibitem[{{Rauscher} \& {Marcy}(2006)}]{2006PASP..118..617R}
{Rauscher}, E. \& {Marcy}, G.~W. 2006, \pasp, 118, 617

\bibitem[{{Saraph}(1970)}]{1970JPhB....3..952S}
{Saraph}, H.~E. 1970, J. Phys. B: At. Mol. Opt. Phys, 3, 952

\bibitem[{{Seaton}(1975)}]{seaton1975}
{Seaton}, M.~J. 1975, Advances Atom. Molec. Phys, 11, 83

\bibitem[{{Shields} {et~al.}(1999){Shields}, {Pogge}, \& {de
  Robertis}}]{1999ASPC..175..353S}
{Shields}, J.~C., {Pogge}, R.~W., \& {de Robertis}, M.~M. 1999, in ASP Conf.
  Ser. 175: Structure and Kinematics of Quasar Broad Line Regions, ed. C.~M.
  {Gaskell}, W.~N. {Brandt}, M.~{Dietrich}, D.~{Dultzin-Hacyan}, \&
  M.~{Eracleous}, 353

\bibitem[{{Smith} \& {Gallagher}(1966)}]{1966PhRv..145...26S}
{Smith}, W.~W. \& {Gallagher}, A. 1966, Physical Review, 145, 26

\bibitem[{{Staanum} {et~al.}(2004){Staanum}, {Jensen}, {Martinussen}, {Voigt},
  \& {Drewsen}}]{2004PhRvA..69c2503S}
{Staanum}, P., {Jensen}, I.~S., {Martinussen}, R.~G., {Voigt}, D., \&
  {Drewsen}, M. 2004, \pra, 69, 032503

\bibitem[{{Taylor} \& {Dunn}(1973)}]{1973PhRvA...8.2304T}
{Taylor}, P.~O. \& {Dunn}, G.~H. 1973, \pra, 8, 2304

\bibitem[{{Theodosiou}(1989)}]{1989PhRvA..39.4880T}
{Theodosiou}, C.~E. 1989, \pra, 39, 4880

\bibitem[{{Vaeck} {et~al.}(1992){Vaeck}, {Godefroid}, \& {Froese
  Fischer}}]{1992PhRvA..46.3704V}
{Vaeck}, N., {Godefroid}, M., \& {Froese Fischer}, C. 1992, \pra, 46, 3704

\bibitem[{{Waller}(1926)}]{waller1926}
{Waller}, I. 1926, Z. f. Phys, 38, 635

\bibitem[{{Wiese} {et~al.}(1969){Wiese}, {Smith}, \&
  {Miles}}]{1969atp..book.....W}
{Wiese}, W.~L., {Smith}, M.~W., \& {Miles}, B.~M. 1969, {Atomic transition
  probabilities. Vol. 2: Sodium through Calcium. A critical data compilation}
  (NSRDS-NBS, Washington, D.C.: US Department of Commerce, National Bureau of
  Standards, |c 1969)

\bibitem[{{Zapesochny{\v i}} {et~al.}(1975){Zapesochny{\v i}}, {Kel'Man},
  {Imre}, {Dashchenko}, \& {Danch}}]{1975SJETP..42..989Z}
{Zapesochny{\v i}}, I.~P., {Kel'Man}, V.~A., {Imre}, A.~I., {Dashchenko},
  A.~I., \& {Danch}, F.~F. 1975, Soviet Physics JETP, 42, 989

\bibitem[{{Zatsarinny} {et~al.}(1991){Zatsarinny}, {Lengyel}, \&
  {Masalovich}}]{1991PhRvA..44.7343Z}
{Zatsarinny}, O.~I., {Lengyel}, V.~I., \& {Masalovich}, E.~A. 1991, \pra, 44,
  7343

\bibitem[{{Zeippen}(1990)}]{1990AA...229..248Z}
{Zeippen}, C.~J. 1990, \aap, 229, 248

\bibitem[{{Zeippen} {et~al.}(1977){Zeippen}, {Seaton}, \&
  {Morton}}]{1977MNRAS.181..527Z}
{Zeippen}, C.~J., {Seaton}, M.~J., \& {Morton}, D.~C. 1977, \mnras, 181, 527

\end{thebibliography}

\end{document}